\documentstyle[aps,prl,twocolumn,psfig,floats]{revtex}
\begin{document}
\draft
\title{Percolation-dependent Reaction Rates in the Etching of Disordered Solids}
\author{K. M. Kolwankar $^{1,2}$, M. Plapp $^2$\
and B. Sapoval $^{1,2}$}

\address{$^1$ Centre de Math\'ematiques et de leurs 
Applications, Ecole Normale Sup\'erieure - CNRS, 
94140 Cachan, France \\
$^2$ Laboratoire de Physique de la Mati\`ere
Condens\'ee, Ecole Polytechnique - CNRS, 
91128 Palaiseau, France}

%\address{}

\maketitle

%\newpage

%{Version \today}

\begin{abstract}

A prototype statistical model for the etching of a random solid
is investigated in order to assess
the influence of disorder and temperature on the 
dissolution kinetics.
At low temperature, the kinetics is dominated by percolation
phenomena, and the percolation threshold determines the global 
reaction time. At high temperature, the fluctuations of the 
reaction rate are Gaussian, whereas at low temperature 
they exhibit a power law tail due to chemical avalanches.
This is an example where microscopic disorder directly induces
non-classical chemical kinetics.

\end{abstract}

\pacs{Pacs: 64.60.Ak, 82.20.-w, 61.43.-j, 05.65.+b}

Chemical etching of disordered solids is a process of great practical 
importance~\cite{EU}. In some cases, such as in corrosion,
it can be a cause of undesirable deterioration. In other 
cases it is useful, for example when one wishes to 
roughen a smooth surface for adhesion or to obtain
some specific optical or esthetic properties. 
Such processes are also stimulating from a theoretical
point of view since it has been observed in 
experiments~\cite{Balazs,claycomb,LD}
that they can give rise to complex patterns and 
dynamics.

To understand the universal features of random etching processes,
the study of simplified toy models is useful since
they allow to grasp some essential results using the 
concepts of percolation theory~\cite{Stauffer}.
The randomness of the solid may be caused either by the topology
of the chemical bonds, such as in amorphous media,
or by fluctuations of the composition, such as in a
solid solution. It can also be the randomness of a 
passivation layer that may appear spontaneously in the
corrosion of polycrystalline solids.

In this letter, we focus on the time evolution of the 
etching process and the influence of temperature. 
In the previous studies of random etching, random
site {\em resistances}  were assigned, and 
``strong" sites were never etched~\cite{model,SS,GBS}.
This has given 
a theoretical interpretation to the spontaneous stabilization 
of fractal surfaces in a model where the etchant is 
consumed in the course of the reaction~\cite{model,SS,GBS}. 
The fluctuations observed in~\cite{Balazs} were interpreted 
in terms of a self-organized fractal growth linked to percolation 
theory~\cite{SS}.

However, if the disorder is at some microscopic 
level, it is more realistic to assign random {\em reaction 
rates} to the dispersed random elements of the solid.
Then, even a site with a very small dissolution rate 
will eventually be etched. Furthermore, 
since real reaction rates are temperature-dependent,
the influence of this parameter
can be studied. In the following, we find a transition from
smooth and continuous growth at high temperatures
to a highly intermittent regime at low temperatures
that exhibits chemical avalanches.

The 2d square lattice model of the solid consists of elements $i$
that have a random site-dependent binding energy $E_i$.
The dissolution rate $R_i$ of a molecule $i$ in contact with
the etchant is supposed to follow an Arrhenius law
that corresponds to an activated process:
\begin{equation}
R_i = C\;\omega_0\; \exp({-E_i/k_BT}) \;,
\label{rates}
\end{equation}
where $C$ is the concentration of the etchant, $k_B$ is
Boltzmann's constant, and $T$ is the temperature.
For simplicity, the attempt frequency $\omega_0$ is 
assumed to be the same for all sites. The energies 
$E_i$ are supposed to be distributed over a finite range.
The simplest choice is $E_i = E_0 + p_i\Delta E$ 
where the $p_i$'s are distributed uniformly between 
0 and 1. Then, Eq.~\ref{rates} can be rewritten as
\begin{equation}
R_i = C\;\omega\; \exp({-{E_\beta} p_i}) \;,
\label{rates1}
\end{equation}
where ${\omega} = {\omega}_0 \exp({-E_0/k_BT})$ and
$E_\beta = \Delta E/k_BT$.
Note that this simplified model neglects 
many factors that may affect real dissolution rates,
such as for example steric blocking.
There are essentially two limits depending on 
the value of $E_\beta$. With $k_BT$ of order 25 meV at
room temperature, weak disorder (``high temperature'')
typically corresponds to $E_\beta$ of order unity, 
and strong chemical disorder (``low temperature'') 
corresponds to $E_\beta$ of order $10$ to $100$.

We consider an irreversible reaction with
no redeposition, and we suppose that
one molecule of etchant is consumed per dissolved site.
It follows that for a finite volume of etchant with initially
$N$ etchant molecules and concentration $C_0$, the concentration 
after the dissolution of $j$ sites is $C_j = C_0(1-j/N)$.
It therefore decreases with time,
and the reaction stops when all the 
etchant molecules are exhausted. Furthermore, we
suppose that the mixing in the solvent (by diffusion 
of the etchant or hydrodynamic flow) is
fast compared to the chemical events, such that 
the concentration in the solution remains uniform.

%---------------------figure------------------------
\begin{figure}
\psfig{figure=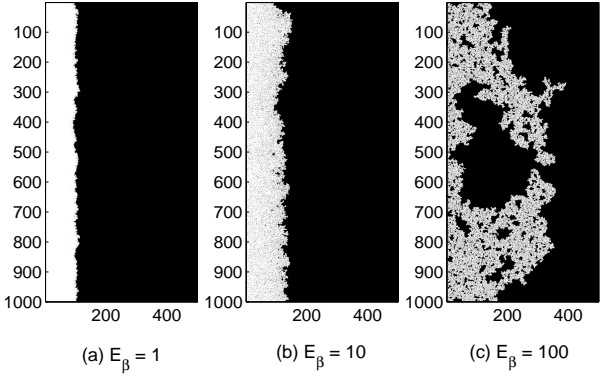,clip=true,width=8cm}
\caption{Snapshot pictures of the etching front at the
end of the reaction for $N=10^5$, $L=10^3$ and (a) $E_\beta=1$,
(b) $E_\beta=10$ and (c) $E_\beta=100$. The unetched solid 
is black and the solution is white. Note that the surface of the 
white regions is the same and equal to N, 
the initial number of etchant molecules in the solution.}
\label{fig1}
\end{figure}
%---------------------------------------------------
The time evolution of etching is studied through an 
event-oriented Monte Carlo algorithm in physical time.
A site $i$ dissolves within a time 
interval $\Delta t$ with probability $R_i\Delta t$
if $\Delta t$ is small enough ($R_i\Delta t \ll 1$).
Since these probabilities are independent for different 
sites, the probability that {\em any} of the surface 
sites is etched in the time interval $\Delta t$ is
$\sum_{i\in \cal S} R_i \Delta t$, where the sum runs over all 
the surface particles (active sites). The time interval
before a site is etched with probability one is therefore
\begin{equation}
\Delta t = {1\over\sum_{i\in \cal S} R_i}
= {1\over C\omega {\int_0^1 h(p)  \exp(-E_{\beta}p) dp}},
\label{tstep}
\end{equation}
where $h(p)$ gives the number of active sites with resistance
between $p$ and $p+dp$. Within that time interval, a given 
site $i$ is etched with probability
\begin{equation}
P_i = {R_i\over \sum_{i\in \cal S} R_i}
= {\exp({-{E_\beta} p_i})\over {\int_0^1 h(p) \exp(-E_{\beta}p) dp}}.
\label{probas}
\end{equation}
In each Monte Carlo step, the algorithm chooses to etch a 
surface site $i$ with probability $P_i$ and increases
the time $t$ by $\Delta t$. As a consequence, the 
physical time is {\it not proportional} to the 
number of Monte Carlo steps, since $\Delta t$ depends on the 
dissolution rates of all the surface sites. However, by 
construction, the average time needed  to dissolve a {\em given} 
site with reaction rate $R_i$ is always the same, 
independently of the system size or the surface configuration. The 
probability distribution of Eq.~\ref{probas} has already been used in 
Ref.~\cite{GCP} to study the influence of finite temperature on invasion 
percolation, but this study was restricted to the initial stages
of the process.

The initial lattice of width $L$ has a flat interface. The 
ratio $N/L$ is the etching depth if uniform 
dissolution occurs. We study here the case $N/L\gg 1$ such that deep pores 
can be formed. In this situation, the essential parameter that governs
the pattern formation is the reduced energy $E_\beta$ 
which controls the relative dissolution speed of ``hard'' and
``weak'' sites. For small $E_\beta$, 
the dissolution rate is nearly the same for all sites 
whatever their energy; in the limit $E_\beta \rightarrow 0$, 
this model maps onto the Eden model \cite{Barabasi}. In this regime,
etching leads to the formation of a microscopically rough, 
but macroscopically smooth interface
that has a finite width $W$ and propagates with a roughly
constant speed (Fig.~\ref{fig1}a). When $E_\beta$ increases, 
sites with small $p_i$ become relatively much faster 
than sites with large $p_i$. The dynamics leads to a
irregular or porous interface (Figs.~\ref{fig1}b and ~\ref{fig1}c),
with disconnected islands surrounded by sites which 
may survive for a long time. These islands continue to 
be slowly etched and their size therefore
decreases with time. In the limit $E_\beta \rightarrow \infty$, 
the weakest site is chosen with probability 1 and the 
model becomes identical to classical invasion percolation ~\cite{WW}.

%--------------------figure-------------------------
\begin{figure}
\psfig{figure=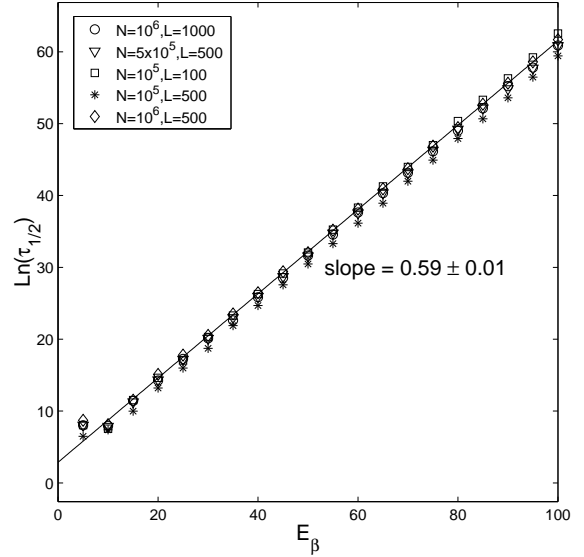,clip=true,width=8cm}
\caption{Activation plot of the reaction half-time $\tau_{1/2}$.} 
\label{fighalflife}
\end{figure}
%---------------------------------------------------
For this reason, it is not surprising that the 
percolation properties of the underlying lattice 
enter the dynamics in
the large $E_\beta$ limit. Our main focus here is
the chemical kinetics. The global reaction rate 
strongly depends on temperature. To characterize 
its behavior, we show in Fig.~\ref{fighalflife} the 
reaction half-time $\tau_{1/2}$, that is, the time it 
takes to consume half of the N etchant molecules, as a function of $E_\beta$. 
For large enough $E_\beta$, this plot gives an effectice 
activation energy equal to $(0.59 \pm 0.01)$, close to 
the square lattice percolation threshold $p_c$. 
The time for the reaction to go from N/2 to N/4
etchant molecules follows the same behavior.
Simulations for the 3D cubic lattice give
a similar result with a slope $0.30 \pm 0.01$,
close to $p_c$ for this 3D lattice.
This indicates that, for strong disorder (or low temperature) 
the global reaction has an activation energy equal to 
$E_0 + p_c\Delta E$ (the $E_0$ contribution comes from the 
definition of $\omega$). 
To our knowledge, this is the first direct appearance 
of a percolation threshold in the {\it time domain}.

\begin{figure}
\psfig{figure=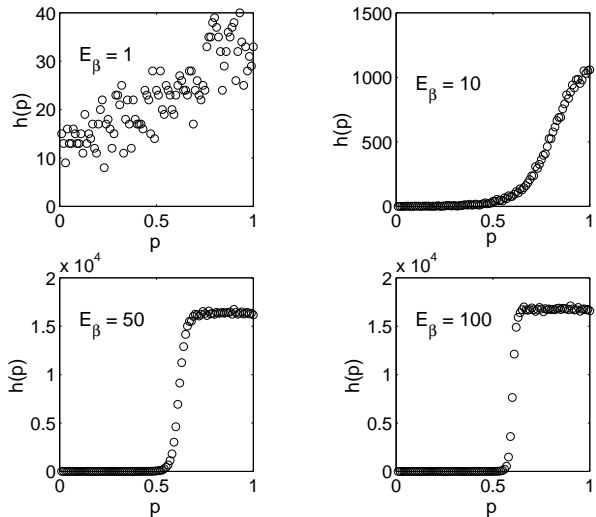,clip=true,width=8cm}
\noindent
\caption{Histogram of $p_i$s on the interface 
after 50\% of the etchant is consumed 
(averaged over 10 realisations). 
Here, $N=10^6$, $L=10^3$.}
\label{fighist}
\end{figure}

Figure~\ref{fighist} shows the histograms $h(p)$
of the site resistances $p_i$ on the solid-liquid 
interface (including the islands) after 50\% of 
the etchant is consumed 
for different temperatures.
They count the number
of surface sites with $p<p_i<p+\epsilon$ as a 
function of $p$ (here $\epsilon=0.01$).
For small $E_\beta=1$ (high temperature),
$h(p)$ increases with $p$, but there is always a large
number of weak sites on the surface. In contrast,
for $E_\beta \gg 1$ (low temperature), there are
very few weak sites, and $h(p)$ starts to increase
sharply at a value close to the percolation threshold.

%----------------------figure-----------------------
%---------------------------------------------------

These data help to explain the appearance of
the percolation threshold in the reaction rate.
The total time needed to etch $n$ molecules is simply
the sum of the time-steps for each individual etching event,
$t = \sum_{j=1}^n \Delta t_j$. At high temperature, all the dissolution rates 
are of similar magnitude. Therefore,
the time-step depends only weakly on the surface
configuration and the reaction proceeds with
a well-defined average rate. In contrast, at
low temperature, the contributions of successive
configurations are very different. For a typical 
configuration that contains a few weak sites, the sum 
in the denominator of Eq.~(\ref{tstep}) is dominated by
the large rates associated with these sites, the 
corresponding time-step is small, and the weakest site is
etched with a probability close to one. However, from
time to time, the growth front coincides with the
perimeter of a percolation cluster. In this 
situation, the weakest site has a strength 
close to $p_c$, and the corresponding time-step 
scales as $\Delta t \sim \exp(E_\beta p_c)$.
As a consequence, the dynamics consists of series 
of rapid etching events (``chemical avalanches'')
separated by long time steps when the
front has reached the surface of a percolation
cluster. The appearance of the percolation
threshold in the reaction time indicates
that the latter are the rate-limiting steps.

\begin{figure}
\psfig{figure=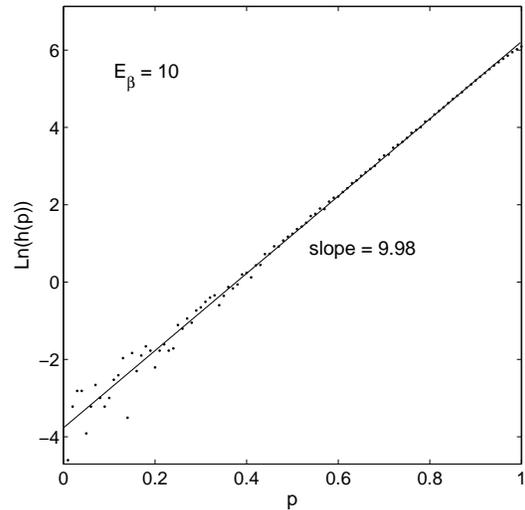,clip=true,width=7cm}
\noindent
\caption{Histogram of $p_i$s on the {\it final} interface
for $E_\beta = 10$, $N=10^6$ and $L=10^3$.}
\label{fighistb10}
\end{figure}

The time evolution of the histograms can be 
qualitatively understood on a coarse-grained 
time scale, that is, a time scale much larger 
than individual time-steps. On this scale, one can 
define a global reaction rate $R$, that is, the 
total number of sites etched per unit time.
Note that $R$ must be proportional to $C\omega$. 
One can then write a continuous evolution equation 
for the number of surface sites of strength $p$:
\begin{equation}
{dh(p)\over dt} = R \epsilon - h(p) C\omega \exp(-E_\beta p) .
\end{equation}
The first term counts the arrival of freshly
uncovered sites (with uniform distribution) on
the surface, whereas the second term is the product 
of the etching rate of sites with strength $p$ times
their number. One clearly sees that 
$h(p) = R\epsilon \exp(E_\beta p)/(\omega C)$ is a 
steady-state solution of this equation. This is indeed 
what is observed for high temperatures. For low 
temperatures, inserting $R\sim C\omega \exp(-E_\beta p_c)$
yields $h(p)\sim \epsilon \exp[E_\beta(p-p_c)]$ \cite{note},
in remarkable agreement with the {\em final} histogram
data for $E_\beta=10$, shown in Fig.~\ref{fighistb10}. 
This has a general consequence that could be called
the ``chemical survival of the fittest'': since almost 
all weak sites have been removed from the surface, the
disordered solid is stronger once etched than it was before
the attack. This has already been discussed in a simpler 
context ~\cite{GBS}.
%----------------------figure-----------------------
%--------------------------------------------------

One notes that, for even lower temperatures, 
the {\em transitory} $h(p)$ shown in Figure~\ref{fighist} levels
off for large $p$. This occurs because in our simulations
the total number of steps is far too small to reach 
the steady state at low temperature. Indeed, for such 
a steady state to appear, a sufficient contrast must occur 
between sites above $p_c$. For this to happen, a huge number 
of strong sites must be uncovered, and this requires 
an exceedingly long time. At zero temperature, which 
is equivalent to invasion percolation, the steady state is never 
reached. The plateau for the highest $p$ values in the transitory 
histograms simply reflects the initial uniform distribution 
of site strengths.

The chemical avalanches that occur at low temperatures 
can be characterized in two different ways. The first is 
purely geometric. When the ``hard'' perimeter of a 
percolation cluster is broken, the reaction has a large 
probability to continue at the place where the
last etching event has occurred since new weak sites
may be uncovered. The avalanche  size can then be determined by 
``connected cluster events''. Starting 
with a given etching event, if the
next etched site is a first neighbor, it is added
to the cluster, and so on, until the next etched
site is not a neighbor to any of the cluster sites.
For large values of $E_\beta$, the number of avalanches
versus their size follows a power law with a slope 
around $1.53 \pm 0.01$ (not shown here), 
in agreement with the result for classical invasion
percolation~\cite{Roux}.

%---------------------figure------------------------
\begin{figure}
\psfig{figure=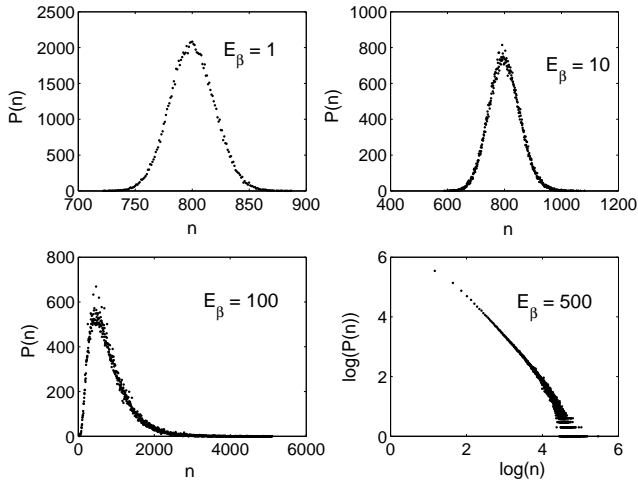,clip=true,width=8.5cm}
\noindent
\caption{Statistical distribution of the reaction rate
for different temperatures: distribution of the 
number $n$ of sites etched in a time interval 
$t_2-t_1=T_{tot}/1000$. 
Here $N=10^6$, $L=10^3$. The first 20\% of events
have been discarded.}
\label{figglrn} 
\end{figure}
%---------------------------------------------------
A more physical way to characterize avalanches is
via the global reaction rate. It can be written as
\begin{equation}
R = {n(t_2-t_1)\over t_2-t_1},
\end{equation}
where $n(t_2-t_1)$ is the number of sites etched
between times $t_1$ and $t_2$. So defined,
this quantity is proportional to an experimental
signal, for example the corrosion current if the
etching is the result of an electrochemical process.
To study the behavior of the reaction rate, we have
performed simulations with a constant etchant
concentration $C$ (corresponding to an infinite
volume of solution) up to a fixed number of etched sites $N$.
The total reaction time $T_{tot}$ was divided in equal
time intervals $t_2-t_1= T_{tot}/1000$, and the number 
of sites etched in each interval was counted 
(discarding the initial etching of the flat 
surface). The resulting histograms are shown in
Fig.~\ref{figglrn}. For small $E_{\beta}$ values, 
the reaction rate has a Gaussian distribution
around a well-defined average.
In contrast, for very
large values of $E_{\beta}$, the histogram becomes
close to a power law due to the occurence
of avalanches with a broad size distribution.
Therefore, the reaction rate has 
large fluctuations in both space and time.
Recently, chemical avalanches have been
found in experiments~\cite{claycomb} and were
explained in terms of a sandpile model~\cite{BTW}.
Our findings might offer an alternative interpretation
for these observations.

In summary, we have introduced a minimal model for 
the etching of a disordered solid. We have studied the influence
of temperature on chemical kinetics and etching 
patterns. At high temperature (or weak disorder), the reaction
rate exhibits gaussian fluctuations around a
well-defined average rate, and the dissolution 
front is macroscopically smooth. At low 
temperature (or strong disorder), the reaction rate has 
strong fluctuations caused by chemical avalanches. 
The global reaction rate is shown to depend on 
temperature with an activation energy directly 
related to the percolation threshold. Interestingly, 
the avalanches, which may be large at low temperature, 
do not influence the reaction rate itself. 
The effect of possible redeposition and that 
of concentration gradients on reaction kinetics 
and patterns should be examined 
in the future.


\begin{references}
\bibitem{note}
This can also be deduced as a fixed point solution of a
reccurrence relation relating the histogram at the $n$th
Monte Carlo step to that of $(n+1)$th.
\end{references}
\end{document}